\documentclass[10pt,letterpaper]{article}
\usepackage[top=0.85in,footskip=0.75in,marginparwidth=2in]{geometry}

\usepackage[utf8]{inputenc}

\usepackage{cite}

\usepackage{nameref,hyperref}

\usepackage[right]{lineno}

\usepackage{microtype}
\DisableLigatures[f]{encoding = *, family = * }

\raggedright
\usepackage{changepage}

\usepackage[aboveskip=1pt,labelfont=bf,labelsep=period,singlelinecheck=off]{caption}

\makeatletter
\renewcommand{\@biblabel}[1]{\quad#1.}
\makeatother

\usepackage{lastpage,fancyhdr,graphicx}
\usepackage{epstopdf}
\fancyheadoffset[L]{2.25in}
\fancyfootoffset[L]{2.25in}

\usepackage{color}

\definecolor{Gray}{gray}{.25}

\usepackage{graphicx}

\usepackage{sidecap}

\usepackage{wrapfig}
\usepackage[pscoord]{eso-pic}
\usepackage[fulladjust]{marginnote}
\reversemarginpar

\usepackage{amsmath,amssymb,amsfonts}
\usepackage{algorithmic}
\usepackage{pgfplots}
\pgfplotsset{compat=1.17} 
\usepackage{pgf-pie}
\usepackage{graphicx}
\usepackage{tikz}
\usepackage{textcomp}
\def\BibTeX{{\rm B\kern-.05em{\sc i\kern-.025em b}\kern-.08em
    T\kern-.1667em\lower.7ex\hbox{E}\kern-.125emX}}

\usepackage{color}
\usepackage{colortbl}
\usepackage{makecell}
\usepackage{tabularx}
\usepackage{pgfplotstable}
\pgfplotsset{compat=newest}
\usepackage{siunitx}
\usepackage{multirow}
\usepackage{pgfplots}
\pgfplotsset{/pgf/number format/use comma} 
\usepgfplotslibrary{statistics}
\usepackage{amsmath,bm}

\newcommand{\HiRel}{  \cellcolor{gray!50} HR}
\newcommand{\Rel}{    \cellcolor{gray!30} R}
\newcommand{\SomeRel}{\cellcolor{gray!10} SR}

\definecolor{Diagramm1}{HTML}{00AC2A}
\definecolor{Diagramm2}{HTML}{63DE17}
\definecolor{Diagramm3}{HTML}{CFFFB2}

\begin{document}
\vspace*{0.35in}

\begin{flushleft}
{\Large
\textbf\newline{Relevance-Based Compression of Cataract Surgery Videos}
}
\newline
\\
Natalia Mathá\textsuperscript{1},
*Klaus Schoeffmann\textsuperscript{1},
Konstantin Schekotihin\textsuperscript{1},
Stephanie Sarny\textsuperscript{2},
Doris Putzgruber-Adamitsch\textsuperscript{2},
Yosuf El-Shabrawi\textsuperscript{2,3}
\\
\bigskip
\bf{1} Klagenfurt University, Klagenfurt, Austria
\\
\bf{2} Klinikum Klagenfurt, Klagenfurt, Austria 
\\
\bf{3} Medical University Graz, Graz, Austria
\\
\bigskip
* Klaus.Schoeffmann@aau.at

\end{flushleft}

\section*{Abstract}
In the last decade, the need for storing videos from cataract surgery has increased significantly. Hospitals continue to improve their imaging and recording devices (e.g., microscopes and cameras used in  microscopic surgery, such as ophthalmology) to enhance their post-surgical processing efficiency. The video recordings enable a lot of user-cases after the actual surgery, for example, teaching, documentation, and forensics.  However, videos recorded from operations are typically stored in the internal archive without any domain-specific compression, leading to a massive storage space consumption. In this work, we propose a relevance-based compression scheme for videos from cataract surgery, which is based on content specifics of particular cataract surgery phases. We evaluate our compression scheme with three state-of-the-art video codecs, namely H.264/AVC, H.265/HEVC, and AV1, and ask medical experts to evaluate the visual quality of encoded videos. Our results show significant savings, in particular up to 95.94\% when using H.264/AVC, up to 98.71\% when using H.265/HEVC, and up to 98.82\% when using AV1.


\section{Introduction}
\textit{Cataract} is a severe clouding disease of the natural eye lens, which often comes with aging and can eventually lead to blindness if not treated. Cataract surgery is a viable treatment option, where an ophthalmic surgeon replaces the natural lens with an artificial one, by using a microscope and tiny instruments. Cataract surgery is the most frequent surgery in the world, and it follows a standardized procedure consisting of 12 phases.     

However, cataract surgery is also an incredibly challenging operation that requires specialized and intensive training. It is crucial to train aspiring surgeons to handle the various aspects and potential complications in the surgical workflow. The microscope that is used for cataract surgery is typically equipped with two eyepieces and allows only one additional person to closely follow the operation in real-time. Although it is possible for a limited number of additional trainees to be present in the operation room (OR), they can only follow the surgery indirectly via the video display that shows a live image of the microscope. 
The display is typically used by clinical personell to follow the surgical workflow and prepare operation equippment. Due to the limited space in the OR, the indirect viewing angle to the display, and the passive/non-interactive nature of the live image, it is not a good source for teaching either. 

Therefore, more and more clinicians record videos of entire surgeries, and use them later for teaching and training via interactive video demonstrations. The video recordings are also suitable for video documentation, which is often a necessity for surgeries and a good alternative to traditional textual reports. The videos can communicate every little detail of a surgery, and even reveal causes for complications, much better than a text report would do. Also, they can be used for forensics and post-operative studies that are performed over a large number of patients. 

Since cataract surgery is the most frequent operation, full video recording and archival can quickly lead to an immense amount of data volume, which causes data management problems for the hospital information systems. Public cloud storage is no solution because clinicians want to keep surgical videos local for various reasons, for example, privacy and existing safety policies. This leads to the very unfortunate situation, that nowadays recorded cataract videos are often deleted after a specific amount of time and the invaluable information is lost and cannot be used for any further purpose.  

The research aim of this work, which is an extension of our previous work \cite{my_compression_cbms}, is to optimize the required storage space of cataract surgery videos, and thereby allow for a long-term archival strategy. We focus on regular cataract surgery videos that consist of they typical 12 phases with no complications or extra phases. Since this is the most frequent situation, our results are applicable to the majority of cataract surgery videos.

In \cite{my_compression_cbms}, we have already defined the relevance of each phase in a regular cataract surgery by a user study with $30$ clinicians and proposed compression parameters for H.264/AVC~\cite{H264AVC} and AV1~\cite{AV1}. In this work we extend this study to H.265/HEVC\cite{H265} and perform additional runtime evaluations. These are the most commonly used codecs and supported by a wide set of web browsers and operating systems. We also evaluate two different compression approaches: (1) considering idle phases as a part of the subsequent phase, and (2) removing them entirely as irrelevant content, as agreed with medical experts. We finally perform a qualitative study with cataract experts, who are requested to inspect and rate the encoded video segments, in order to verify if the produced visual quality is still acceptable for all surgical phases. 
Our results show that regular cataract surgery phases can be significantly compressed without loss of medically relevant content.


\section{Related Work}
Many papers have been published in the literature, which focus on improving video compression. Most of them are specific video coding schemes, or concepts to improve the general compression of video data, for example \cite{NewEncoderExample} and \cite{ImprovedEncoderExample}. Some recent works also utilize neural networks for the purpose of general video coding, e.g., \cite{NewEncoderCNN2},~\cite{NewEncoderCNN3}, and~\cite{NewEncoderCNN1}.
Another research area is the utilization of intra-frame information in videos. For example, the papers~\cite{AttentionCompression2},~\cite{AttentionCompression1},~\cite{AttentionCompression3} compress videos by encoding less relevant content in every frame with lower quality, where the relevance is defined using an attention mechanisms.

However, few research has been conducted on relevance-based video compression in specific domains, such as medicine. The challenge here is the automatic assessment of relevance in the content, which follows the same rules a medical expert would apply. Hence, the works that consider clinicians' subjective evaluations are essential, as they can identify the compression parameters that match the end-user expectations.

In~\cite{BronchoscopicCompression}, the authors evaluate the compression efficiency of bronchoscopic surgery videos. Their method encodes bronchoscopic recordings using different compression rates and video codecs, such as M-JPEG 2000~\cite{mjpeg2000} and  H.264/AVC~\cite{H264AVC}. A hybrid vector measure metric~\cite{HVMmetric} and Hosaka plots~\cite{HosakaPlots} are used to evaluate each configuration. The results achieve a compression rate of $96$ (of the uncompressed content). 

In~\cite{BerndLaporoscopyCompression}, the authors perform a user study to assess the possible trade-off between low bitrate and medically satisfying visual quality of laparoscopy videos. For this study, they encode laparoscopic videos with the H.264/AVC~\cite{H264AVC}) codec and different encoding parameters, such as resolution and constant rate factor. As a result of the user study, the authors identify an acceptable compression rate and finally propose three configurations for \textit{visually lossless} quality, \textit{good quality}, and \textit{acceptable quality}. Using the proposed recommendations, the authors can save additional $60\%$, $87.5\%$, and $92.86\%$ of required storage space, respectively (when compared to common H.264/AVC encoding).

For the field of cataract surgery, \cite{NeginRelevance} proposes a relevance-based compression scheme using intra-frame relevance. First, the authors encode idle phases in cataract surgery videos (with no surgical tools inside the eye) with low quality, because such idle phases are considered irrelevant. Next, they apply region-of-interest (ROI) detection to the \textit{cornea} area as well as instruments, and give less priority to the other spatial content in the frame. As a result, the proposed method allows to save up to $68\%$ of storage space using content-specific compression and removal of irrelevant content.

Although some other research works focused on surgical video compression (as described above), to the best of our knowledge, there is no work that considers the medical content relevance in cataract surgery videos, as defined by clinicians. 
We propose a novel approach to compress cataract surgery videos based on the content's relevance, as defined by surgeons.
We manually extract the phase samples for each relevance category and compress them with different visual parameters. Finally, we let clinicians evaluate visual quality of the result segments in a user study, to find the acceptable compression rate.

\section{Clinical Relevance of Phases in Cataract Surgery}
\label{sec:clinicalrelevance}

The procedure of cataract surgery is pretty standardized. According to medical experts~\cite{my_generalization}, each common cataract surgery consists of $12$ phases, distinguishable by the used instruments. During these phases, the surgeon performs tasks with varying difficulty. For example, filling up the patient's eye with antibiotics or a viscoelastic substance is rather simple, while removing the natural lens or inserting an artificial one requires very specific skills. Each surgical phase is followed by an idle phase, where the surgical instruments are changed. Different phases of cataract surgery have different relevance, as rated by the clinicians.

In \cite{my_compression_cbms} we have presented results of a user study conducted with $30$ medical experts who defined the clinical relevance of different content segments in regular cataract surgery videos for different purpose (teaching, documentation, and research). They rated the relevance with different levels: (N: \textit{not relevant at all}, SR: \textit{somewhat relevant}, R: \textit{relevant}, and HR: \textit{highly relevant}). The results are summarized again in Table~\ref{tab:relevance_results}. 
 According to the findings, teaching is considered the most crucial aspect for clinicians, as it received the highest relevance score among all purposes surveyed for each phase. 
For relevance-based video compression this means that a higher video quality/bitrate is needed for teaching (with less quality degradation), while for documentation and research purpose the content can be compressed more efficiently (i.e., more strongly quantized, with less remaining visual quality).

\begin{table}[!h]
\caption{Relevance rates of regular cataract surgery phases for different purpose: teaching, documentation, and research, as determined by medical experts (median value). HR, R, and SR are abbreviations for \textit{Highly Relevant}, \textit{Relevant}, and \textit{Somewhat Relevant}.}
\label{tab:relevance_results}
    \centering
    \begin{tabular}{|r|ccc|}
         \hline 
         \textbf{Surgery Phase}                        &  \textbf{Teaching}    & \textbf{Documentation}  & \textbf{Research} \\ \hline \hline
         Incision                                   & \Rel                  & \SomeRel              & \SomeRel          \\ \hline
         Viscoelastic I                             & \SomeRel              & \SomeRel              & \SomeRel          \\ \hline
         Capsulorhexis                              & \HiRel                & \Rel                  & \Rel              \\ \hline
         Hydrodissection                            & \HiRel                & \SomeRel              & \SomeRel          \\ \hline
         Phaco
                                                    & \HiRel                & \Rel                  & \Rel              \\ \hline
         \makecell[r]{Irrigation/\\aspiration}      & \Rel                  & \SomeRel              & \SomeRel          \\ \hline
         \makecell[r]{Capsule \\polishing}            & \SomeRel              & \SomeRel              & \SomeRel          \\ \hline
         Viscoelastic II                            & \SomeRel              & \SomeRel              & \SomeRel          \\ \hline
         Implantation                               & \Rel                  & \Rel                  & \Rel              \\ \hline
         \makecell[r]{Viscoelastic \\aspiration}    & \Rel                  & \SomeRel              & \SomeRel          \\ \hline
         \makecell[r]{Sealing of \\incisions}       & \Rel                  & \SomeRel              & \SomeRel          \\ \hline
         \makecell[r]{Antibiotic \\injection}       & \Rel                  & \Rel                  & \Rel              \\ \hline 
    \end{tabular}
\end{table}

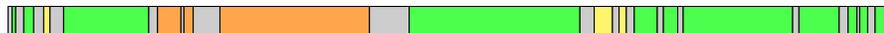
\begin{figure*}[!ht]
    \centering
    \pgfplotsset{testbar/.style={
        xbar stacked, area style,
        width=\textwidth,
        axis y line*= none, axis x line*= none,
        y axis line style={opacity=0},
        x axis line style={opacity=0},
        xmajorgrids = false,
        xmin=0,xmax=100,
        ytick = \empty,
        xtick = \empty,
        yticklabels = {},
        xticklabels = {},
        tick align = outside, xtick pos = left,
        bar width=4mm, y=5mm,
        enlarge y limits={abs=0.625},
        }}
    \begin{tikzpicture}
        \begin{axis}[testbar,legend style={area legend, at={(0.5,-0.85)}, anchor=north, legend columns=-1}] 
            \addplot[fill=gray!40] coordinates{(0.45,0)};
            
            \addplot[fill=green!70] coordinates{(0.39,0) }; 
            \addplot[fill=gray!40] coordinates{(0.92,0)};
            
            \addplot[fill=green!70] coordinates{(1.12,0)}; 
            \addplot[fill=gray!40] coordinates{(1.14,0)};
            
            \addplot[fill=yellow!70] coordinates{(0.71,0) }; 
            \addplot[fill=gray!40] coordinates{(1.54,0)};
            
            \addplot[fill=green!70] coordinates{(9.62,0) }; 
            \addplot[fill=gray!40] coordinates{(1.00,0)};
            
            \addplot[fill=orange!70] coordinates{(2.65,0) }; 
            \addplot[fill=gray!40] coordinates{(0.33,0)};
            
            \addplot[fill=orange!70] coordinates{(1.06,0) }; 
            \addplot[fill=gray!40] coordinates{(3.03,0)};
            
            \addplot[fill=orange!70] coordinates{(16.90,0) }; 
            \addplot[fill=gray!40] coordinates{(4.50,0)};
            
            \addplot[fill=green!70] coordinates{(19.3,0) }; 
            \addplot[fill=gray!40] coordinates{(1.63,0)};
            
            \addplot[fill=yellow!70] coordinates{(2.05,0) }; 
            \addplot[fill=gray!40] coordinates{(0.76,0)};
            
            \addplot[fill=yellow!70] coordinates{(0.81,0) }; 
            \addplot[fill=gray!40] coordinates{(0.90,0)};
            
            \addplot[fill=green!70] coordinates{(2.60,0) }; 
            \addplot[fill=gray!40] coordinates{(0.69,0)};
            
            \addplot[fill=green!70] coordinates{(1.64,0) }; 
            \addplot[fill=gray!40] coordinates{(0.60,0)};
            
            \addplot[fill=green!70] coordinates{(12.38,0) }; 
            \addplot[fill=gray!40] coordinates{(0.72,0)};
            
            \addplot[fill=green!70] coordinates{(4.55,0) }; 
            \addplot[fill=gray!40] coordinates{(1.01,0)};
            
            \addplot[fill=green!70] coordinates{(1.01,0) }; 
            \addplot[fill=gray!40] coordinates{(0.31,0)};
            
            \addplot[fill=green!70] coordinates{(0.90,0) }; 
            \addplot[fill=gray!40] coordinates{(0.83,0)};
            
            \addplot[fill=green!70] coordinates{(1.09,0) }; 
            \addplot[fill=gray!40] coordinates{(0.60,0)};
        \end{axis}
        \end{tikzpicture}
    \caption{Example of a cataract surgery video: temporal relevance distribution for teaching purpose. The content relevance is given by the block color: idle phases (\textit{not relevant}) are gray, \textit{somewhat relevant} content is yellow, \textit{relevant} segments are green, and \textit{highly relevant} content is orange.}
    \label{fig:surgery_representation}
\end{figure*}

Fig.~\ref{fig:surgery_representation} depicts an illustration of an exemplary cataract surgery video, along with all phases and the corresponding relevance rates obtained for teaching purpose. A relatively long part of cataract surgery (e.g., in this case $75.48\%$) has \textit{relevant} and \textit{highly relevant} content.

\section{Relevance-Based Cataract Video Compression}
\label{sec:materials_compr}

To evaluate the achievable compression gain, we use the relevance-rates defined in the previous section and encode videos of different phases in cataract surgery with different encoding settings (codecs, bitrates, resolutions). 

\begin{figure}[!h]
    \centering
    \includegraphics[width=3.25in]{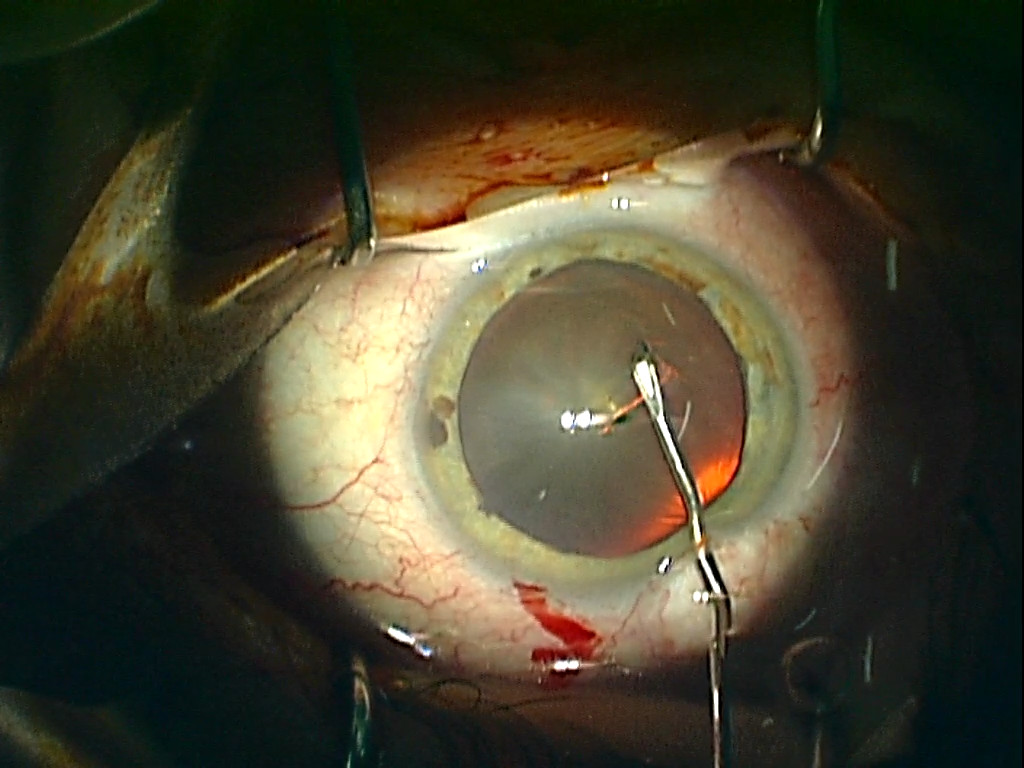}
    \caption{Exemplary video frame of the \textit{capsulorhexis} phase.}
    \label{fig:example_caps}
\end{figure}

\begin{figure}[!h]
    \centering
    \includegraphics[width=3.25in]{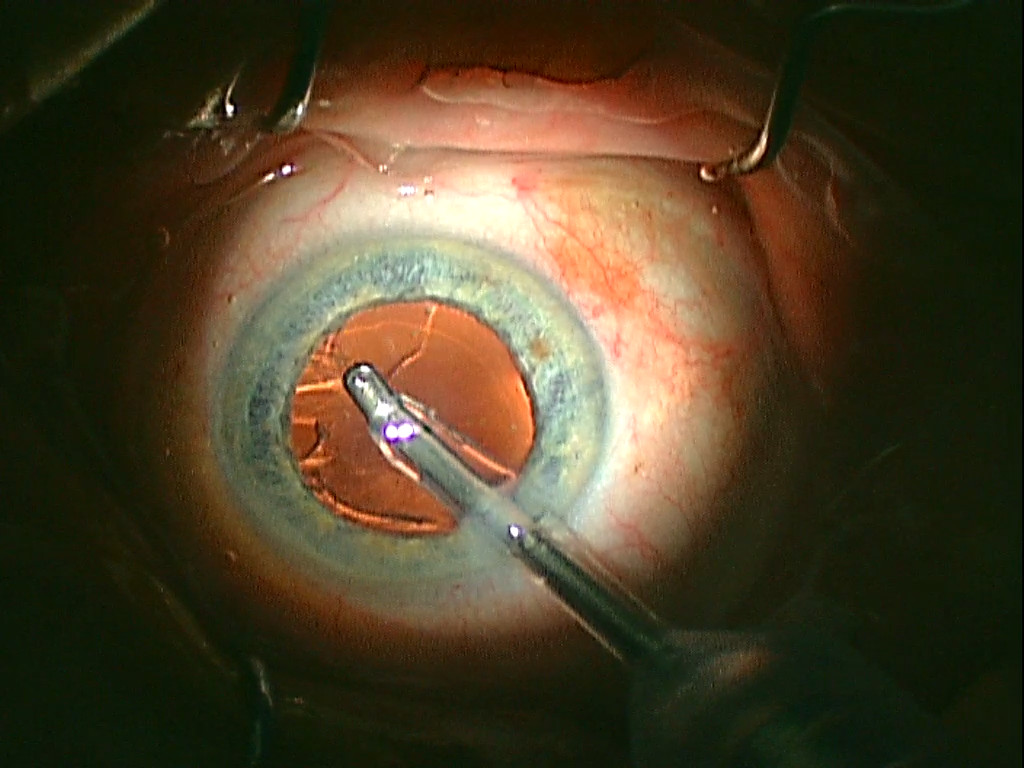}
    \caption{Exemplary video frame of the \textit{irrigation/aspiration} phase.}
    \label{fig:example_IA}
\end{figure}

The underlying video dataset, out of which clips are selected, was recorded at Klinikum Klagenfurt and uses a resolution of $1024x768$ pixels. Example frames are presented in Fig.~\ref{fig:example_caps} and Fig~\ref{fig:example_IA} for the phases \textit{capsulorhexis} and \textit{irrigation/aspiration}. 
It is important to understand that with the current common settings in the hospital (H.264/AVC codec, 60 fps, a constant rate factor (CRF) of 14-16) a typical cataract surgery video would result in a file size of $506$ MiB. This is the basis to which we compare our relevance-based compression method that uses different visual quality for different surgery phases, according to the relevance levels defined by clinicians (see Section~\ref{sec:clinicalrelevance}).

\subsection{Compression Setup}

We assign the highest possible relevance level based on the ratings obtained from the relevance detection survey for each cataract surgery phase (see Table~\ref{tab:relevance_results}). If a phase is \textit{highly relevant} for teaching and \textit{relevant} for research purposes, we categorize it as \textit{highly relevant}. Furthermore, we select the following compression setups (using ffmpeg~\cite{FFMPEG}) for clips from cataract surgery videos: 

    \begin{itemize}
        \item H.264/AVC, $23$-$47$ CRF (with a step of $2$);
        \item H.265/HEVC, $23$-$47$ CRF (with a step of $2$);
        \item AV1, $27$-$63$ CRF (with a step of $3$);
    \end{itemize}

The constant-rate-factor (CRF) is ffmpeg's way to control the visual quality (i.e., encoding bitrate) in an inverse setting (a CRF value of $0$ produces the best quality).
For H.264/AVC and H.265/HEVC the lower value ($47$) is chosen because otherwise the resulting videos have unacceptable visual quality, i.e., very pixelated, while the upper CRF is a default value ($23$).
For AV1, the lower CRF value ($63$) is the lowest possible number provided by ffmpeg. To achieve the same number of setups for each codec, we set the upper CRF value ($27$) three points higher than the standard value ($23$). 
We also select different resolution settings, namely 1024x768 (original resolution), 800x600, and 640x480 pixels. 

It becomes clear that due to the high number of parameters (13 bitrate settings, 12 phases, three resolutions), we would end up with $13x12x3 = 468$ different clips/settings to test for each codec in the qualitative study, which is practically impossible. Hence, to maintain the participants' focus while still evaluating a large number of different encoding configurations, we select one representative clip from each group of relevance categories (HR, R, and SR) only. More specifically, to reveal the maximum compression rate for cataract surgery phases and to find the optimal compression parameters, we consider \textit{capsulorhexis} as \textit{highly relevant} (original bitrate $12278$ kbps), \textit{irrigation/aspiration} as \textit{relevant} ($12416$ kbps), and \textit{viscoelastic I} and \textit{viscoelastic II} as \textit{somewhat relevant} ($13074$ kbps). Also, we test only meaningful resolution settings and exclude those that we could rule out in a pre-study already. With this configuration, we end up with $78$ setups for H.264/AVC and AV1, and $39$ setups for H.265/HEVC. 

\subsection{User Study}

We evaluate the achievable compression settings by a qualitative study with eight clinicians (cataract surgery experts). 
The breakdown of the subjects' experience in carrying out cataract surgery is as follows: two experts have \textit{$10$ or more years} of surgical experience, one surgeon operates since nine years, two clinicians have an experience of three years, one has two years of practice, and two experts perform the surgeries since one year or less. Half of the participants exclusively work in private hospitals, while the other half work in both private and public sectors. Of all the survey respondents, six have experience in teaching, seven have performed patient documentation, and six have published scientific papers in this field.

To compare the visual quality of the encoded videos, we utilize the SSIM metric \cite{Objective_Metrics_Comparison}, which evaluates the similarity between frames before and after encoding. The highest quality is represented by a value of $1$, indicating that both frames are identical. In essence, the greater the number of artifacts present in the encoded frame, the lower its SSIM value will be.

To conduct the user study, we arrange the videos for each phase based on the average SSIM value of all per-frame SSIM value. In the following we refer to all the configurations by their \textit{setup number}, where $1$ has the best quality, and $78$ has the worst quality for H.264/AVC and AV1. For H.265/HEVC, the best and worst quality values are $1$ and $39$, respectively.

The videos are presented to the users following a Dichotomous Search Method paradigm. More specifically, the subjects first see the video with the middle quality (e.g., $39$) from an initial interval (e.g., $[1;78]$) and decide if this quality is sufficient. If so, the upper boundary is updated with this setup number, otherwise the lower boundary is updated. For each phase, this method converges within around six to seven steps, which helps the participants to stay focused and allows the rating in a logarithmic time.

\subsection{Relevance-Based Compression Results}

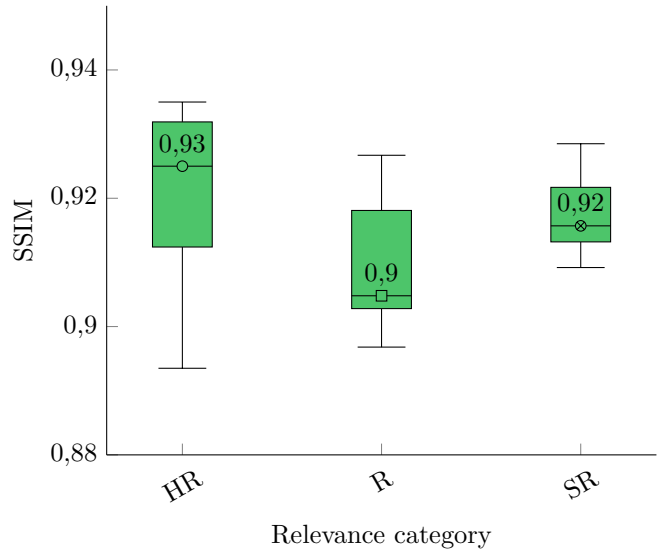
\begin{figure}[!ht]
    \centering
      \begin{tikzpicture}
      \begin{axis}[
      width=3.5in,
      height=2.975in,
      boxplot/draw direction=y,
      axis x line*=bottom,
      axis y line*=left,
      ymin=0.8800,
      ymax=0.9500,
      xtick={1,2,3},
      xticklabels={HR,R,SR,},
      xticklabel style={rotate=30},
      ylabel={SSIM},
      xlabel={Relevance category},
      /pgfplots/boxplot/box extend=0.3,
      boxplotcolor/.style={color=black,fill=#1!70,mark options={color=black,fill=#1!70}}
      ]
      \addplot+[boxplotcolor=Diagramm1,
        boxplot prepared={
         lower whisker=0.8935, lower quartile=0.9124,
         median=0.9250, upper quartile=0.9319,
         upper whisker=0.9350},
         nodes near coords,
         nodes near coords align=south,
        ]
      coordinates {(0,0.9250)};
      \addplot+[boxplotcolor=Diagramm1,
        boxplot prepared={
         lower whisker=0.8968, lower quartile=0.9028,
         median=0.9048, upper quartile=0.9181,
         upper whisker=0.9267},
         nodes near coords,
         nodes near coords align=south,
        ]
      coordinates {(0,0.9048)};
      \addplot+[boxplotcolor=Diagramm1,
        boxplot prepared={
         lower whisker=0.9092, lower quartile=0.9132,
         median=0.9157, upper quartile=0.9217,
         upper whisker=0.9285},
         nodes near coords,
         nodes near coords align=south,
        ]
      coordinates {(0,0.9157)};
      \end{axis}
      \end{tikzpicture}
    \caption{Evaluation with H.264/AVC and AV1. Experts' rates distribution with the median values on the acceptable quality for different relevance groups, where \textit{HR}, \textit{R}, and \textit{SR} stand for \textit{Highly Relevant}, \textit{Relevant}, and \textit{somewhat relevant}, respectively.}
    \label{fig:experts_opinion_h264_av1}
\end{figure}

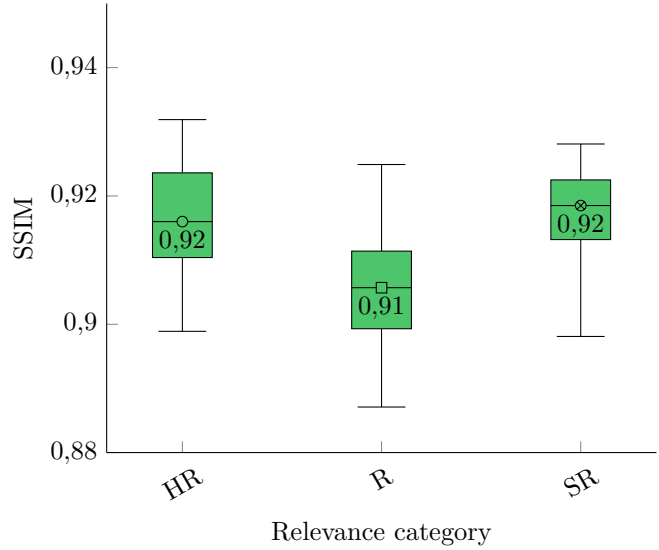
\begin{figure}[!ht]
    \centering
      \begin{tikzpicture}
      \begin{axis}[
      width=3.5in,
      height=2.975in,
      boxplot/draw direction=y,
      axis x line*=bottom,
      axis y line*=left,
      ymin=0.8800,
      ymax=0.9500,
      xtick={1,2,3},
      xticklabels={HR,R,SR,},
      xticklabel style={rotate=30},
      ylabel={SSIM},
      xlabel={Relevance category},
      /pgfplots/boxplot/box extend=0.3,
      boxplotcolor/.style={color=black,fill=#1!70,mark options={color=black,fill=#1!70}}
      ]
      \addplot+[boxplotcolor=Diagramm1,
        boxplot prepared={
         lower whisker=0.8989, lower quartile=0.9104,
         median=0.9160, upper quartile=0.9236,
         upper whisker=0.9319},
         nodes near coords,
         nodes near coords align=north,
        ]
      coordinates {(0,0.9160)};
      \addplot+[boxplotcolor=Diagramm1,
        boxplot prepared={
         lower whisker=0.8871, lower quartile=0.8993,
         median=0.9057, upper quartile=0.9114,
         upper whisker=0.9249},
         nodes near coords,
         nodes near coords align=north,
        ]
      coordinates {(0,0.9057)};
      \addplot+[boxplotcolor=Diagramm1,
        boxplot prepared={
         lower whisker=0.8981, lower quartile=0.9132,
         median=0.9185, upper quartile=0.9225,
         upper whisker=0.9281},
         nodes near coords,
         nodes near coords align=north,
        ]
      coordinates {(0,0.9185)};
      \end{axis}
      \end{tikzpicture}
    \caption{Evaluation with H.265/HEVC. Experts' rates distribution with the median values on the acceptable quality for different relevance groups, where \textit{HR}, \textit{R}, and \textit{SR} stand for \textit{Highly Relevant}, \textit{Relevant}, and \textit{somewhat relevant}, respectively.}
    \label{fig:experts_opinion_h265}
\end{figure}

Fig.~\ref{fig:experts_opinion_h264_av1} and Fig.~\ref{fig:experts_opinion_h265} present the study results. Please note that the visual quality improves with the increase of SSIM metric. We observe that although the clinicians' results are diverse, the experts require better visual quality for \textit{highly relevant} phases and slightly lower visual quality for \textit{relevant} one.

To determine the ideal compression parameters, we utilize the median value rounded down to an integer for each relevance category (Fig.~\ref{fig:experts_opinion_h264_av1}) as a threshold value for visual quality.
In particular, these median integer values are $36$ ($0.9250$ SSIM) for \textit{highly relevant} content, $59$ ($0.9048$ SSIM) for \textit{relevant}, and $53$ ($0.9157$ SSIM) for \textit{somewhat relevant} category for encoded with H.264/AVC and AV1 videos. For H.265/HEVC, these values are $17$ ($0.9160$ SSIM), $22$ ($0.9057$ SSIM), and $17$ ($0.9185$ SSIM). To select appropriate compression parameters for each relevance group, we drop the options where the SSIM value is lower than the defined threshold. Afterward, we select the configurations with the lowest bitrate for each codec. Table~\ref{tab:compression_results} shows the result setups.

\begin{table*}[!h]
\caption{The optimal compression parameters for H.264/AVC, H.265/HEVC, and AV1 with SSIM and bitrate for each relevance category.}
\label{tab:compression_results}
    \centering
    \begin{tabular}{|c|c|cccccc|}
         \hline 
         \textbf{Codec}             & \textbf{Relevance}    & \textbf{Setup} & \textbf{Resolution}   & \textbf{CRF}  & \textbf{SSIM} & \textbf{kbps}   & \textbf{Compr.}  \\ \hline 
         \hline
         \multirow{3}{*}{H.264/AVC}     & highly                    & 34                & 640x480         & 25            & 0.9260        & 653.39                    & 94.68\%                  \\ 
                                    & relevant                     & 56                & 640x480         & 33            & 0.9079        & 252.11                    & 97.97\%                  \\ 
                                    & somewhat                     & 52                & 640x480         & 31            & 0.9162        & 248.49                    & 98.10\%                  \\ \hline 
         \hline

         \multirow{3}{*}{H.265/HEVC}     & highly                    & 16                & 640x480         & 31            & 0.9174        & {207.12}            & {98.31\%}          \\ 
                                    & relevant                      & 22                & 640x480         & 35            & 0.9057        & {130.10}            & {98.95\%}          \\ 
                                    & somewhat                     & 16                & 640x480         & 31            & 0.9202        & {173.66}            & {98.67\%}          \\ \hline 
        \hline
        
         \multirow{3}{*}{AV1}       & highly                    & 36                & 1024x768        & 57            & 0.9250        & \textbf{190.38}           & \textbf{98.45\%}          \\ 
                                    & relevant                      & 57                & 800x600        & 63            & 0.9078        & \textbf{68.32}            & \textbf{99.45\%}          \\ 
                                    &somewhat                     & 51                & 640x480         & 60            & 0.9179        & \textbf{75.46}            & \textbf{99.42\%}          \\ \hline

    \end{tabular}
\end{table*}

We can see that the bitrates of these compression configurations are relatively low with at least $653.39$ kbps for H.264/AVC, $207.12$ for H.265/HEVC, and $190.38$ kbps for AV1. Besides, we discover a bitrate of just $68.32$ kbps with AV1 as optimal compression parameters for \textit{relevant} content. Using the proposed settings allows optimizing the required storage space for at least $94.68\%$ for H.264/AVC, at least $98.31\%$ for H.265/HEVC, and at least $98.45\%$ for AV1 for the entire surgery considering highly relevant content.

\subsection{Idle Phases}

According to clinicians, idle phases, where no instruments are being used, typically do not include any relevant content. However, as there may be some unforeseen exceptions, which is why we explore two methods of processing idle phases. Each idle phase can be deemed part of the preceding phase and compressed accordingly, or it can be considered a separate video segment. Medical professionals classify idle phases as completely irrelevant content if they are treated as independent segments. As a result, such segments can be entirely eliminated from cataract surgery videos.
Thus, we present an additional assessment of the potential storage savings that could be achieved by removing idle phases. To this end, we gather and annotate $20$ typical cataract surgery videos manually in terms of idle and non-idle phases, frame by frame (for a total of $492,425$ frames). 
The annotations are carried out by a skilled technician under the guidance of medical experts.
It turns out that in all of these videos, idle phases take around $23.75\%$ of space. 
In other words, considering idle phases as irrelevant content, allows to save at least \textbf{$95.94\%$}, \textbf{$98.82\%$}, and \textbf{$98.71\%$} of space for each evaluated codec by removing them from cataract surgery videos before compressing the rest of the video files.

\subsection{Impact of Experience}

We have calculated the correlation between participants' practicing experience and their preferences for encoded video quality. Our findings reveal that there exists a moderate negative correlation between participants' experience and their expectations for visual quality in \textit{highly relevant} and \textit{somewhat relevant} content, specifically $70.20\%$ and $63.47\%$, correspondingly.
At the same time, for the \textit{relevant} phases, it is lower with $40.96\%$. We conclude that with increasing clinicians' experience, the requirements to the visual quality of the recorded surgery decrease.

\subsection{Run-Time Requirements}

Finally, we also evaluate the encoding time for each codec with the defined best configurations (see Table~\ref{tab:compression_results}), which we measure for two different devices, namely:
\begin{enumerate}
    \item A desktop computer with the following parameters:
    \begin{itemize}
        \item Intel(R) Core(TM) i5-10210U CPU @ 1.60GHz 2.11 GHz;
        \item 8 GB RAM;
        \item Windows 10;
        \item ffmpeg version 5.1;
    \end{itemize}
    \item An AWS instance with the following parameters:
    \begin{itemize}
        \item c5.2xlarge instance type:
        \begin{itemize}
            \item 8 vCPUs;
            \item 16 GB RAM;
            \item Intel Xeon Platinum 8124M physical processor;
            \item 3 GHz clock speed;
        \end{itemize}
        \item Windows Server Base 2022;
        \item ffmpeg version 5.1;
        \item $0.756$ USD per hour.
    \end{itemize}
\end{enumerate}

Table~\ref{tab:speed_measurements} shows the obtained results. The values are the average of  several measurements. As expected, we can see that the fastest encoding is reached with the H.264/AVC codec. The AV1 codec requires more than $10$ times encoding time and only saves $1.32$\%-$3.77$\% of storage space, when compared to H.264/AVC. At the same time, H.265/HEVC requires only twice as much time as H.264/AVC but also saves $0.57$\%-$3.63$\%. We can conclude that H.265/HEVC provides the best trade-off solution between the encoding speed and stored space.

\begin{table}[!h]
\caption{The encoding speed measurements for H.264/AVC, H.265/HEVC and AV1 for each relevance category. \textit{HR}, \textit{R}, and \textit{SR} stand for \textit{Highly Relevant}, \textit{Relevant}, and \textit{somewhat relevant}, respectively.}
\label{tab:speed_measurements}
    \centering
    \begin{tabular}{|c|c|ccc|}
         \hline 
         \textbf{Device}    & \textbf{Codec} & \textbf{HR} & \textbf{R}     & \textbf{SR}   \\ \hline
         \hline
         \multirow{3}{*}{Desktop PC}& H.264/AVC          & $\bm{12.1}$s & $\bm{4.3}$s    & $\bm{1.2}$s   \\
                            & H.265/HEVC          & $25.6$s     & $8.4$s         & $2.3$s        \\
                            & AV1            & $140.3$s    & $29.4$s        & $7.1$s        \\ \hline
        \hline
         \multirow{3}{*}{AWS}
         \multirow{3}{*}{Instance}& H.264/AVC          & $\bm{7.7}$s  & $\bm{2.9}$s & $\bm{1.2}$s     \\
                            & H.265/HEVC          & $15.5$       & $5.2$s      & $2.1$s          \\
                            & AV1            & $103.5$      & $27.5$s     & $8.1$s          \\ \hline
    \end{tabular}
\end{table}

\section{Conclusion}

In this paper, we performed several evaluations related to the storage space requirements of videos from cataract surgery. The aim was to optimize the storage space while still keeping the necessary clinical visual quality by using stronger compression (more quantization) for less relevant content, but keeping the clinically necessary visual quality. Our evaluations show that it is possible to save $94.68\%$ of storage space for \textit{highly relevant content} when encoded with H.264/AVC ($98.31\%$ and $98.45\%$ when using H.265/HEVC and AV1). Also, for \textit{relevant} content (which is the vast majority in a video of cataract surgery) we can save $97.97\%$ with H.264/AVC ($95.85\%$ and $99.45\%$ with H.265/HEVC and AV1). Finally, for \textit{somewhat relevant} content it allows to save $98.10\%$ with H.264/AVC ($95.67\%$ and $99.42\%$ with H.265/HEVC and AV1). These savings can be achieved while keeping most of the original video quality (more than $0.9057$ SSIM), which is sufficient for teaching purpose at least. While the saved storage space is a little higher with H.265/HEVC and AV1, our evaluations also show that both codecs require significantly more encoding time (H.265/HEVC about 2x and AV1 about 10x). This may be a problem for hospitals with many procedures, because either more efficient hardware is needed, or a continuously increasing backlog of video encoding jobs would build up. Hence, we conclude that the H.265/HEVC codec offers the best compromise in terms of fast encoding speed and high storage space savings.

Our findings also showed that clinicians prioritize superior visual quality for \textit{highly relevant} content and that there is a moderate negative correlation between age/experience and the need for visual quality for \textit{highly relevant} and \textit{somewhat relevant} content. From this, we can infer that more experienced clinicians tend to have less stringent demands for visual quality.

\section*{Acknowledgment}
This work was funded by the FWF Austrian Science Fund under grant P 31486-N31 and by the EFRE, REACT-EU and Carinthian Economic Promotion Fund Programme (Project ONTIS, Contract No. KWF-3520|34826|50900).

\nolinenumbers

\bibliography{main}

\bibliographystyle{abbrv}

\end{document}